\input harvmac
\input epsf
%\draftmode

\Title{\vbox{\rightline{EFI-2000-9}\rightline{hep-th/0003101}}}
{\vbox{\centerline{On the relevance of tachyons}}}
\vskip10pt

\baselineskip=12pt
\centerline{Jeffrey A. Harvey\footnote{$^a$}{harvey@theory.uchicago.edu}, 
David Kutasov\footnote{$^b$}{kutasov@theory.uchicago.edu}
{\it and} 
Emil J. Martinec\footnote{$^c$}{e-martinec@uchicago.edu}} 
\medskip
\centerline{\sl Enrico Fermi Inst. and Dept. of Physics}
\centerline{\sl University of Chicago}
\centerline{\sl 5640 S. Ellis Ave., Chicago, IL 60637, USA}

\baselineskip=16pt
 
\vskip 2cm
\noindent

We study condensation of open string tachyons using 
renormalization group flow in the worldsheet field theory.
This approach leads to a simple picture of the
physics of the nontrivial condensate.

\Date{3/2000}

%%%%%%%%%%%%%%%%%%%%%%%%%%%%%%%%%%%%%%%%%%%%%%%%%%%%%%%%%%%%%%%%%%%%%%%%
%%%%%%%%%%%%%%%%%%%%%%%%%%%%%%%%%%%%%%%%%%%%%%%%%%%%%%%%%%%%%%%%%%%%%%%%
%	Emil Martinec's macros
%
%
\def\bibitem{\nref}

\def\ie{{\it i.e.}}

\def\cf{{\it c.f.}}

\def\sst{\scriptscriptstyle}

\def\frac#1#2{{#1\over#2}}
\def\coeff#1#2{{\textstyle{#1\over #2}}}
\def\half{\frac12}

\def\ket#1{|#1\rangle}

\def\d{\partial}

\def\inbar{\,\vrule height1.5ex width.4pt depth0pt}
\def\IC{\relax\hbox{$\inbar\kern-.3em{\rm C}$}}
\def\IR{\relax{\rm I\kern-.18em R}}
\def\IP{\relax{\rm I\kern-.18em P}}

%
%%%%%%%%%%%%%%%%%%%%%%%%%%%%%%%%%%%%
%

%
\catcode`\@=11
\def\slash#1{\mathord{\mathpalette\c@ncel{#1}}}
\overfullrule=0pt

\def\FF{{\cal F}}

\def\ZZ{{\cal Z}}

\def\underrel#1\over#2{\mathrel{\mathop{\kern\z@#1}\limits_{#2}}}

\catcode`\@=12

%%%%%%%%%%%%%%%%%%%%%%%%%%%%%%%%%%%%%%%%%%%%%%%%%%%%%%%%%%%%%%

%

\def\ket#1{\left| #1\right\rangle}

\def\tr{{\rm tr}}

\def\exp{{\rm exp}}

%%%%%%%%%%%%%%%%%%%%%%%%%%%%%%%%%%%%%%%%%%%%%%%%%%%%%%%%%%%%%%
% new defs:

\def\uv{{\sst \rm UV}}
\def\ir{{\sst \rm IR}}
\def\X{{\bf X}}
\def\str{{\sst \rm str}}
\def\gstr{g_\str}
%%%%%%%%%%%%%%%%%%%%%%%%%%%%%%%%%%%%%%%%%%%%%%%%%%%%%%%%%%%%%%
%%%%%%%%%%%%%%%%%%%%%%%%%%%%%%%%%%%%%%%%%%%%%%%%%%%%%%%%%%%%%%
%	References
%
\bibitem\bh{
K.~Bardakci and M.~B.~Halpern,
``Explicit Spontaneous Breakdown In A Dual Model 1\&2,''
Phys.\ Rev.\  {\bf D10}, 4230 (1974),
%``Explicit Spontaneous Breakdown In A Dual Model. 2. N Point Functions,''
Nucl.\ Phys.\  {\bf B96}, 285 (1975);
K.~Bardakci,
``Spontaneous Symmetry Breakdown In The Standard Dual String Model,''
Nucl.\ Phys.\  {\bf B133}, 297 (1978).}
\bibitem\kossam{
V. A.~Kostelecky and S. Samuel,
``On a Nonperturbative Vacuum for the Open Bosonic String,''
Nucl. Phys. {\bf B336}, 263 (1990).}
\bibitem\gns{
E.~Gava, K.~S.~Narain and M.~H.~Sarmadi,
``On the bound states of p- and (p+2)-branes,''
Nucl.\ Phys.\  {\bf B504}, 214 (1997)
[hep-th/9704006].}
\bibitem\senone{
A.~Sen,
``Tachyon condensation on the brane antibrane system,''
JHEP {\bf 9808}, 012 (1998)
[hep-th/9805170].}
\bibitem\sred{
M.~Srednicki,
``IIB or not IIB,''
JHEP {\bf 9808}, 005 (1998)
[hep-th/9807138].}
\bibitem\sentwo{
A.~Sen,
``SO(32) spinors of type I and other solitons on brane-antibrane pair,''
JHEP {\bf 9809}, 023 (1998)
[hep-th/9808141].}
\bibitem\senthree{
A.~Sen,
``Type I D-particle and its interactions,''
JHEP {\bf 9810}, 021 (1998)
[hep-th/9809111].}
\bibitem\wK{
E.~Witten,
``D-branes and K-theory,''
JHEP {\bf 9812}, 019 (1998)
[hep-th/9810188].}
\bibitem\horava{
P.~Horava,
``Type IIA D-branes, K-theory, and matrix theory,''
Adv.\ Theor.\ Math.\ Phys.\  {\bf 2}, 1373 (1999)
[hep-th/9812135].}
\bibitem\senfour{
A.~Sen,
``BPS D-branes on non-supersymmetric cycles,''
JHEP {\bf 9812}, 021 (1998)
[hep-th/9812031].}
\bibitem\sensix{
A. Sen,
``Non-BPS States and Branes in String Theory,''
[hep-th/9904207].}
\bibitem\senseven{
A.Sen,
``Supersymmetric world-volume action for non-BPS D-branes,''
[hep-th/9909062].}
\bibitem\senfive{
A.~Sen,
``Universality of the tachyon potential,''
JHEP {\bf 9912}, 027 (1999)
[hep-th/9911116].}
\bibitem\sz{
A.~Sen and B.~Zwiebach,
``Tachyon condensation in string field theory,''
[hep-th/9912249].}
\bibitem\brk{N.~Berkovits,
``The Tachyon Potential in Open Neveu-Schwarz String Field Theory,''
[hep-th/0001084].}
\bibitem\wt{
W. ~Taylor,
``D-Brane effective field theory from string field theory,''
[hep-th/0001201].}
\bibitem\hk{
J.~A.~Harvey and P.~Kraus,
``D-branes as unstable lumps in bosonic open string field theory,''
[hep-th/0002117].}
\bibitem\bsz{N.~Berkovits, A. ~Sen and B.~Zwiebach,
``Tachyon Condensation in Superstring Field Theory,''
[hep-th/0002211].}
\bibitem\mt{
N.~Moeller and W.~Taylor,
``Level truncation and the tachyon in open bosonic string field theory,''
[hep-th/0002237].}
\bibitem\kjmr{
R.~de Mello Koch, A.~Jevicki, M. Mihailescu and R.~Tatar,
``Lumps and Branes in Open String Field Theory,''
[hep-th/0003031].}
\bibitem\fsw{
P.~Fendley, H.~Saleur and N.~P.~Warner,
``Exact solution of a massless scalar field 
with a relevant boundary interaction,''
Nucl.\ Phys.\  {\bf B430}, 577 (1994)
[hep-th/9406125].}
\bibitem\al{
I.~Affleck and A.~W.~Ludwig,
``Universal noninteger 'ground state degeneracy' in critical quantum systems,''
Phys.\ Rev.\ Lett.\  {\bf 67}, 161 (1991).}
\bibitem\altwo{
I.~Affleck and A.~W.~Ludwig,      
``Exact conformal field theory results on the multichannel Kondo effect:
single fermion Green's function, self energy, and resistivity,''
Phys.\ Rev.\ {\bf B48}, 7297 (1993).}
\bibitem\hkms{
J.~A.~Harvey, S.~Kachru, G.~Moore and E.~Silverstein,
``Tension is dimension,''
JHEP {\bf 0003}, 001 (2000)
[hep-th/9909072].}
\bibitem\ers{
S.~Elitzur, E.~Rabinovici and G.~Sarkissian,
``On least action D-branes,''
Nucl.\ Phys.\  {\bf B541}, 246 (1999)
[hep-th/9807161].}
\bibitem\cklm{
C.~G.~Callan, I.~R.~Klebanov, A.~W.~Ludwig and J.~M.~Maldacena,
``Exact solution of a boundary conformal field theory,''
Nucl.\ Phys.\  {\bf B422}, 417 (1994)
[hep-th/9402113].}
\bibitem\pt{
J.~Polchinski and L.~Thorlacius,
``Free fermion representation of a boundary conformal field theory,''
Phys.\ Rev.\ {\bf D50}, 622 (1994)
[hep-th/9404008].}
\bibitem\fls{
P.~Fendley, F.~Lesage and H.~Saleur,
``Solving 1-d plasmas and 2-d boundary problems 
using Jack polynomials and functional relations,''
J.\ Stat.\ Phys. {\bf 79}, 799 (1995)
[hep-th/9409176].}
\bibitem\tcsa{
P.~Dorey, A.~Pocklington, R.~Tateo and G.~Watts,
``TBA and TCSA with boundaries and excited states,''
Nucl.\ Phys.\  {\bf B525}, 641 (1998)
[hep-th/9712197];
P.~Dorey, I.~Runkel, R.~Tateo and G.~Watts,
``g-function flow in perturbed boundary conformal field theories,''
[hep-th/9909216].}
\bibitem\mttt{
D.~Mitchell and N.~Turok,
``Statistical Properties Of Cosmic Strings,''
Nucl.\ Phys.\  {\bf B294}, 1138 (1987).}
\bibitem\zam{
A.~B.~Zamolodchikov,
``Conformal Symmetry And Multicritical Points in 
Two-Dimensional Quantum Field Theory,''
Sov.\ J.\ Nucl.\ Phys.\  {\bf 44}, 529 (1986).}
\bibitem\lss{
F.~Lesage, H.~Saleur and P.~Simonetti,
``Boundary flows in minimal models,''
Phys.\ Lett.\  {\bf B427}, 85 (1998)
[hep-th/9802061].}
\bibitem\aff{
I.~Affleck,
``Conformal Field Theory Approach to the Kondo Effect,''
Acta Phys.\ Polon.\  {\bf B26}, 1869 (1995)
[cond-mat/9512099].}
\bibitem\bm{
T.~Banks and E.~Martinec,
``The Renormalization Group And String Field Theory,''
Nucl.\ Phys.\  {\bf B294}, 733 (1987).}
\bibitem\fri{
D. Friedan, unpublished; see also \bm.}
\bibitem\pol{
J.~Polchinski,
``Vertex Operators In The Polyakov Path Integral,''
Nucl.\ Phys.\  {\bf B289}, 465 (1987);
``Factorization Of Bosonic String Amplitudes,''
Nucl.\ Phys.\  {\bf B307}, 61 (1988).}
\bibitem\lpp{
A.~LeClair, M.~E.~Peskin and C.~R.~Preitschopf,
``String Field Theory On The Conformal Plane. 1. Kinematical Principles,''
Nucl.\ Phys.\  {\bf B317}, 411 (1989);
``String Field Theory On The Conformal Plane. 2. Generalized Gluing,''
Nucl.\ Phys.\  {\bf B317}, 464 (1989).}
\bibitem\gm{
P.~Ginsparg and G.~Moore,
``Lectures On 2-D Gravity And 2-D String Theory,''
Lectures given at TASI summer school, Boulder CO, June 11-19, 1992
[hep-th/9304011].}

%%%%%%%%%%%%%%%%%%%%%%%%%%%%%%%%%%%%%%%%%%%%%%%%%%%%%%%%%%%%%%

The configuration space of open string theory is of intrinsic
interest; it is the arena for D-brane dynamics, and may teach us
about other nonperturbative string phenomena. An example of such
a phenomenon that has received some attention recently 
\refs{\bh-\kjmr}
is tachyon condensation in unstable D-brane systems. It has been 
proposed \senone\ that the endpoint of the condensation is the 
closed string vacuum.  

An approach using level truncation of
open string field theory yields remarkably accurate results
for quantities such as the vacuum energy \refs{\kossam,\sz,\brk,\bsz,\mt}
and the properties of low-lying excitations \refs{\hk,\kjmr}
of the nontrivial condensate, which appear to support
these ideas. However, it is not easy to understand
why this should be so within the context of string
field theory. Also, it is difficult to study
the properties of the non-trivial vacuum using this approach. 

In this note, we point out that the worldsheet boundary
renormalization group provides a simple conceptual framework
for understanding open string tachyon condensation.  We will see
that an analysis of 
the RG flows induced by boundary perturbations yields a 
qualitative and quantitative physical picture of the open string
configuration space.  This makes it possible to find non-trivial
classical solutions and study excitations around them.

In particular, using results in the literature one can prove that 
unstable branes indeed annihilate by tachyon condensation, compute 
the energy of the vacuum, and show that all open string modes
disappear from the spectrum in the process. One can also study flows 
in which an unstable brane (or brane-antibrane pair) annihilates into 
any number of lower dimensional branes.

\vskip .5cm
\noindent{\bf The bosonic string}
\medskip
String dynamics is succinctly specified as the condition
of Weyl invariance on the string worldsheet $\Sigma$.  The classical
open string equations of motion are the vanishing of the
beta functions for boundary perturbations of the 
two-dimensional action 
on the disk. 
Flowing to the IR on the worldsheet is equivalent to approaching a
classical solution of the spacetime theory.
Thus the study of tachyon condensation in 
open string theory is equivalent to an examination of the RG flow
induced by boundary perturbations, for example 
\eqn\tach{
  \int_{\d\Sigma}d\tau \;T[X(\tau)]
}
in the open bosonic string, with $T[X]$ slowly varying in spacetime.
  
To begin, consider the boundary RG flow with a single cosine potential
\eqn\bsg{
  T[X_{25}]=-\lambda\cos[kX_{25}]
}
\ie\ the boundary Sine-Gordon model.  We choose conventions
such that $\alpha'=1$. Thus $k^2<1$ for a relevant perturbation;
we restrict our discussion to this situation.
The theory with boundary interaction \tach\
is exactly solvable \fsw\ using Bethe ansatz techniques.  
The UV fixed point is free field theory
with Neumann boundary conditions.  The coupling $\lambda$ 
increases under flow to the IR, pinning the boundary value
of the field to one of the minima of the potential, which
are spaced a distance $\delta X_{25}=2\pi/k$. 
Thus, the IR fixed point is a stack of D24-branes.

Under boundary RG flow, the trace of the stress-energy tensor
is only nonzero on the boundary; the bulk theory remains
conformal, and in particular the central charge $c$ is fixed.
Thus the flow is in the space of boundary conditions of
a given bulk theory.  It is believed that there is a quantity 
analogous to $c$, which measures the `number of boundary degrees 
of freedom', and decreases along RG flows.  This quantity is
the boundary entropy $g$ \al, which can be defined as the term 
in the annulus partition function which is independent of the 
width $L$ of the strip in the thermodynamic limit
$L\rightarrow\infty$
\eqn\partfn{
  \beta\FF=\beta\FF_{\rm bulk}-\log[g]\ ,
}
or equivalently as the disk partition function $g_B=\langle 0|B\rangle$
of the boundary state $\ket B$ associated to the perturbed theory.
Affleck and Ludwig \refs{\al,\altwo} showed that $g$ decreases along 
RG flows at the lowest nontrivial order in conformal perturbation theory,
but there is as yet no proof that this is always the case.

Quite generally, $g$ measures the tension of the corresponding
D-brane in string theory \hkms; physically, the $g$-conjecture
should be related to minimization of the action in the space of
open string fields.

In any event, for the boundary Sine-Gordon interaction one has 
(from the exact solution using the thermodynamic Bethe ansatz \fsw)
\eqn\guvgir{
  \frac{g_{\ir}}{g_{\uv}}=k\ .
}
This result agrees with the ratio of energy densities of the IR stack
of D24-branes relative to the UV D25-brane. Indeed, assuming
that $X_{25}$ is compactified on a large circle of radius $R$,
the boundary entropy is\foot{A rescaling is needed to relate
$g_\uv$ and $g_\ir$ to the tensions of the corresponding
branes \hkms.}
\refs{\fsw,\ers}:
\eqn\giruv{\eqalign{
g_{\uv}&= \sqrt{R}\cr
g_{\ir}&= kR \cdot \frac{1}{\sqrt{R}} \cr
}}
where $kR$ is the number of D24-branes at the IR fixed point
of the flow \bsg. $g_{\ir}/g_{\uv}$ computed from \giruv\ agrees with 
the exact calculation \guvgir; it is also consistent with the conjectured
``$g$-theorem'', since $g_{\ir}/g_{\uv}< 1$ precisely when the perturbation 
\bsg\ is relevant.  

In the case $k^2=1$ the perturbation \bsg\ is truly marginal
and the coupling $\lambda$ parametrizes a line of fixed points,
interpolating between Neumann (small $\lambda$) and Dirichlet
(large $\lambda$) boundary conditions \refs{\cklm,\pt}. The boundary entropy
$g$ is independent of $\lambda$, as one would expect from the
$g$-conjecture. 

The RG flow of the interaction \tach\ has also been studied
using conformal perturbation theory \fls.
While TBA techniques establish the exact result for
the IR boundary entropy \guvgir, it is useful to compare
the conformal perturbation theory with the method of level
truncation in string field theory \refs{\kossam,\sz}.  The latter appears
to converge quite rapidly.  Conformal perturbation theory
is simply the expansion of the partition function in a
power series in $\lambda$
\eqn\cpt{
  \ZZ=\sum_{n=0}^{\infty}\lambda^{2n}\ZZ_{2n}\ ,
}
where the coefficients $\ZZ_{2n}$ are given by the
integrated correlation functions of $2n$ of the tachyon
perturbations \tach.  Bounds on the $\ZZ_{2n}$ establish
that the radius of convergence of the expansion is infinite
for $k^2<\half$ \fls. Expanding the boundary entropy
$\log[g]=-\d\FF/\d T$ in $\lambda$, the series converges quite 
rapidly; for example, $\log[g_\ir]$ is obtained to within 4\%
for $k^2=\coeff13$, keeping only the first seven terms in the
expansion \fls.  Thus conformal perturbation theory
provides a viable alternative to the level truncation
approach in string field theory as 
an approximation to the dynamics.\foot{Another 
approximation scheme, which appears to be quite close in spirit
to level truncation of the string field theory, is the 
so-called truncated conformal space approach \tcsa.
In a conformal field theory perturbed by 
a boundary operator, one diagonalizes the perturbed Hamiltonian
in the finite dimensional space of open string states 
having less than a given energy in the unperturbed theory.  
One again obtains remarkably
good agreement with exact results using spaces on the order
of 100 states.}

The limit $k\rightarrow 0$ isolates a single D24-brane,
if we remain at a finite point in space; alternatively,
if we perturb by ${\sin}(kX_{25})$, the branes
are pushed off to infinity as $k\rightarrow 0$,
and we are left with the closed string vacuum as
conjectured in \senone.  In the limit, the term in the
energy proportional to the volume vanishes, 
since $g_\ir/ g_\uv\rightarrow 0$.

One might wonder what has happened to the dynamics of open strings
as a result of the tachyon condensation.  From the worldsheet point
of view, all boundary operators $\exp(ipX_{25})$ are flowing
to the identity operator under the RG flow (they become
c-numbers when the boundary conditions on $X_{25}$ are Dirichlet);
$X_{25}$ becomes localized at one of the minima of the potential \bsg.  
In spacetime, normalizable eigenstates of the Hamiltonian 
in the unstable vacuum $T=0$ (corresponding to the UV fixed point of \bsg) 
are wavepackets which move freely in $X^{25}$;
at the IR fixed point, all the normalizable modes are bound to the
D24-brane, as was found for the lowest modes in \hk. 
In addition, since the kinetic terms for these spacetime fields are
given by the $L_0$ eigenvalue of the corresponding worldsheet
operator, these terms vanish in the IR as suggested in
\senseven.

The worldsheet analysis is also consistent with the discussion of ref. 
\kossam, where evidence was presented that after the tachyon condenses, 
the inverse propagator of some low lying fields no longer has any 
zeroes, which means that they do not give rise to physical
excitations (in the context of our present discussion, the analysis
of \kossam\ describes the region far from the D24-branes). The worldsheet
picture shows that this is true for all modes of the open string, as one
would expect.

It is important to emphasize that this process
of binding of physical excitations to a wall is entirely
classical in string theory. A breakdown of the classical approximation
would manifest itself as a singularity in the RG flow corresponding
to \tach, \bsg. Such a singularity certainly does not exist for
the flow \bsg\ and there is no reason to expect one for the more
general case \tach. Note also that the above discussion is true 
in particular for the open string gauge field, and for any fields 
that couple to it.  The disappearance of the gauge field dynamics 
in the bulk is thus a {\it classical} phenomenon.

At finite $k$, a finite spacing between D24-branes is maintained
at the IR fixed point.  There should be open string sectors
corresponding to the stretched strings between different branes;
how do these arise?  At any point along the flow, the cosine
potential will have solitons on the boundary, where the
boundary value of $X_{25}$ hops from one minimum of the cosine
to another.  The soliton creation and annihilation operators
will flow to the vertex operators that create and destroy
the corresponding stretched strings at the IR fixed point.
These solitons descend from highly excited open string states.
In the typical such state, the configuration of the
open string is a random walk \mttt; the cosine potential
then draws the endpoints to the minima at $X_{25}=2\pi n/k$,
but not necessarily the same one at both endpoints.
The lightest stretched strings presumably descend from the leading
Regge trajectory $(\d X_{25})^N$ of the UV fixed point, 
which generates a straight stretched string 
of length $\sim\sqrt N$ in string units.

We next turn to the case of a general slowly varying tachyon condensate
$T(X_{25})$ depending only on $X_{25}$. Generically, $T$
will have a series of minima at some points $X_i$, near which it behaves
like $T(X)\sim (X-X_i)^2$. The infrared fixed point of \tach\ will in this
case describe a series of D24-branes located at the points $X_i$. One can
ask what happens when different minima coincide and one considers a
multicritical potential $T(X)\sim (X-\hat X)^{2m}$, near some point $\hat X$.
If $T(X)$ were a bulk perturbation, the system would flow in the IR to
the $(m+1, m+2)$ minimal model, with central charge $c=1-\frac6{(m+1)(m+2)}$ 
\zam. The multicritical boundary potential appears instead to describe 
in the IR a free CFT with $U(m)$ Chan-Paton factors.

As different $X_i$ approach each other, strings stretched between
different minima of $T$ go to zero mass (or worldsheet scaling dimension),
and in the limit, one finds an $m^2$-fold degeneracy of all operators.
A priori, it is not obvious that the sole effect of this degeneracy is
to introduce Chan-Paton factors -- the system could in principle be 
interacting in the IR, but the picture of coalescing D-branes suggests 
that it is in fact free. 
%
%The above discussion is consistent with \lss, where it was shown
%that the ratio of the boundary entropies resulting from a flow 
%$X^{2m}\rightarrow X^{2m-2}$ in the boundary interaction is
%\eqn\minmod{
%  \frac{g_\ir}{g_\uv}=\frac{m-1}m\ .
%}
There is some evidence \lss\ 
that the multicritical boundary theories with 
potential $T(X)\sim X^{2m}$ are related to the underscreened Kondo 
model with spin $(m-1)/2$, which has the same boundary entropy \aff.
This relation might help in making the connection with Chan-Paton dynamics,
but we are not going to pursue this issue here.

{}From the structure of the tachyon potential in open string 
field theory, $V(T)\sim -\half T^2+bT^3$, (valid for small $T$)
it might appear that there is a dramatic difference in 
the dynamics for opposite signs of $T$.  However, from the 
worldsheet perspective, it is clear that perturbations by 
$\pm T$ lead to qualitatively the same physics. As discussed 
above, turning on a perturbation \tach\ leads in the IR to 
physics dominated by the minima of the potential $T(X)$. If we 
take $T\to -T$, we find similar physics, with the degrees of 
freedom localized at the maxima of $T$. 

For example, taking $X$ to be compact and turning on a 
slowly varying tachyon field, which is an arbitrary 
combination of relevant boundary operators, one finds
both for $T$ and for $-T$, a maximum number 
$[R]$ of D24-branes 
which can move around in $X^{25}$ and annihilate. Of 
course, in agreement with perturbative string theory 
there is no $T\to - T$ symmetry, as the properties of 
the minima and maxima of $T$, for a generic given $T$, 
are not the same. 

The apparent instability of the spacetime potential $V(T)$ 
for negative $T$ is reflected in the worldsheet approach
as the fact that if $T(X)$ (for non-compact $X$) is slowly
varying but not bounded from above, $-T$ is not bounded from 
below, and the worldsheet theory appears to be sick. From the 
general perspective it is clear that one should think of this
instability as describing a D24-brane (that is a minimum of $T$)
that was pushed to $|X|\to\infty$. 

So far we discussed tachyon profiles that depend only on $X_{25}$.
The D24-branes obtained in the infrared clearly have a tachyon on
their worldvolume which can condense further to make lower D-branes.
The general tachyon condensate depending on all twenty-five spatial
coordinates leaves some set of isolated D0-branes; Dp-branes are 
simply metastable fixed points of the flows. 

One can also study time-dependent tachyon profiles. To use results 
from boundary field theory, one should presumably
Wick-rotate to Euclidean time
and repeat the previous discussion. As before, in the IR the worldsheet
dynamics is localized at a particular value of $X_0$. This value
might be interpreted as the time of annihilation of the 
unstable D-brane.

\vskip .5cm
\noindent{\bf The fermionic string}
\medskip
In the type II superstring, there are unstable non-BPS $Dp$ branes
with even $p$ for type IIB and odd $p$ for type IIA (for a review, 
see \sensix). Sen has shown \refs{\sentwo,\senthree}
that the stable, BPS $D(p-1)$ brane can be described as 
a tachyonic kink soliton solution of the spacetime effective
field theory.  
We can again describe this situation via a perturbation of
the worldsheet field theory by a boundary superpotential
\eqn\superact{
  S_{\rm bdy}=\int_{\d\Sigma}d\tau d\theta\left[\Gamma D\Gamma
	+\Gamma\;T[\X_p]\right]\ .
}
Here, $\X_p=x_p+\theta\psi_p$ is an N=1 worldsheet superfield; 
$D=\partial_\theta+\theta\partial_\tau$ is the usual superderivative;
and $\Gamma=\eta+\theta F$ is
an anticommuting quantum-mechanical degree of freedom
living on the boundary, which is needed in order that 
the tachyon vertex operator describes a spacetime boson \wK.
Equivalently, one can add $2 \times 2$ Chan-Paton factors to 
the vertex \sensix.  
Integrating over the boundary supercoordinate $\theta$,
and eliminating the auxiliary field $F$, one finds a bosonic
potential $T^2$, and a boundary fermion interaction 
$\eta \psi_pT'(x_p)$.
In this case, zeroes of the tachyon profile
$T$ contribute supersymmetric minima of the worldsheet boundary,
leading to codimension one branes.

In the IR, the potential acts as a mass term that decouples $\eta$,
and turns the boundary condition on $\X_p$ from Neumann to 
Dirichlet.  One cannot write a tachyon perturbation of the
IR fixed point because the extra boundary fermion $\eta$
needed to write it is absent (similarly for all the other
GSO-odd vertex operators); thus the infrared theory
at a minimum of the boundary potential is GSO projected.
Zeroes of $T$ with opposite signs of $T'$ give opposite contributions
to $\tr(-1)^F$ in the boundary quantum mechanics,
so the corresponding boundary states will have opposite
signs in the contribution of the RR sector (periodic fermion 
boundary conditions in the closed string channel are equivalent
to inserting $\tr(-1)^F$ in the open string channel).
In other words, the minima give (say) $D(p-1)$ branes
(boundary state $\ket{NSNS}+\ket{RR}$),
while the maxima yield $\overline{D}(p-1)$ branes
(boundary state $\ket{NSNS}-\ket{RR}$).
Equivalently, the RR charge of the kink arises due to an
$\int C_p\wedge dT$ coupling in the spacetime effective action.
The sign of the brane charges reflects the sign of $T'$ 
at the zeroes of $T$.

One can generalize the discussion of \guvgir\ to this case by turning
on a superpotential $T=\lambda\cos(k\X_p)$. There is no analog of the exact
solution \fsw\ for this case, but one can still compute the change in
the boundary entropy using the brane picture and check consistency
with the $g$-conjecture.

The tensions of the relevant branes are
\eqn\tensions{\eqalign{
  \mu_p^{\sst\rm (un)} &=\frac{\sqrt 2}{\gstr}\cdot(2\pi)^{-p}\cr
  \mu_{p-1}^{\sst\rm (st)} &=\frac{1}{\gstr}\cdot(2\pi)^{-(p-1)}
}}
where $\mu_p^{\sst\rm (un)}$ and 
$\mu_{p-1}^{\sst\rm (st)}$ are the tensions of the
unstable $p$-brane and stable (BPS) $(p-1)$-brane, respectively.
The spacing of the alternating stack of $D(p-1)$ and $\overline{D}(p-1)$
branes is $\half\cdot(2\pi/k)$, so
\eqn\superguvgir{
  \frac{g_\ir}{g_\uv}=\frac{2kR\;
	\mu_{p-1}^{\sst\rm (st)}}{2\pi R\;\mu_p^{\sst\rm (un)}}=\sqrt 2 \;k\ .
}
Relevance of the superpotential $T$ implies that $k^2<1/2$ in this case;
hence, the result \superguvgir\ is again consistent with the 
$g$-conjecture.

The disappearance of the kinetics in the $p^{\rm th}$ direction
is the same as in the bosonic string.  
The worldsheet analysis shows that all the open string
fields are bound to a lower-dimensional brane after
the tachyon condenses.  There are no normalizable excitations
of the open string fields in the bulk.
The effect is again classical.

It is easy to generalize the above analysis to the boundary flow
corresponding to the generation of a non-BPS $D(p-1)$ brane
as a tachyonic kink on a nearby $Dp-\overline{D}p$ pair. The
calculation \superguvgir\ works in a similar fashion. One finds 
a non-GSO projected spectrum in the IR in this case
as a consequence of the fact that the analog of $\eta$ \superact\
is complex in the $Dp-\overline{D}p$ system. One combination
of $\eta$, $\eta^*$ becomes massive in the flow to the IR, while 
the other remains massless and allows one to construct excitations
with odd worldsheet fermion number like the tachyon on the non-BPS
$(p-1)$-brane. One can also use boundary RG to study the decay of 
non-BPS states on K3 \senfour\  (and other compactifications) as one
varies the closed string moduli of the compactification.

The discussion of a general superpotential $T(X)$ \superact\
is very similar to the bosonic case. One difference is that
there is no longer any instability; the bosonic potential
$T^2$ is always bounded from below. In fact, the RG flow
corresponding to \superact\ has a manifest symmetry,
which takes $T\to -T$
(exchanging BPS $D(p-1)$ and $\overline{D}(p-1)$
branes in the IR limit),
and flips the sign of all the $RR$ fields (thus exchanging
$D$ and $\overline{D}$ branes back). 

\vskip .5cm
\noindent{\bf Discussion}
\medskip
One may wonder what is the precise relation between boundary
RG flows and open string field theory.  In particular, why do
calculations in the level-truncated string field theory 
\refs{\kossam,\sz,\brk,\bsz,\mt}
converge so rapidly to the right answer for the vacuum energy?
We close with some speculations in this regard.

There are strong analogies between string field theory equations 
of motion and the exact (Wilsonian) RG \bm.  One studies the RG 
flows of the theory with a finite cutoff and all possible operators 
in the action. In this interpretation, the Wilsonian RG equations 
are thought of as the Langevin equation of stochastic quantization 
\fri\ of the string field theory. RG scale is stochastic time, in 
which the couplings relax toward an extremum of the action. The 
complete set of couplings includes operators involving the boundary 
curvature and its derivatives; in the `conformal normal ordering' 
prescription of Polchinski \pol, these are encoded in couplings to 
the reparametrization ghosts.  Formally, the space of interacting 
string fields considered in \refs{\senfive,\sz} can be mapped onto 
the space of general boundary perturbations using the formalism of 
\lpp\ combined with Polchinski's prescription.  Thus we expect that 
open string field theory represents a formulation of the exact boundary
RG in a very particular renormalization scheme.

One might then interpret the approximation of adding more and
more levels of the string field as the standard procedure
of RG improvement by adding small amounts of irrelevant operators
to the cutoff theory in order to more closely approximate
the continuum limit.  Typically such a procedure does well 
even when only the first few irrelevant terms are included;
the effects of highly irrelevant operators are highly
damped by their rapid decay.  This may provide a qualitative
explanation of the rapid convergence of the string field
theory calculations of \refs{\kossam,\sz,\mt}.

Another natural set of questions surrounds 
closed string tachyon condensation.  
Following the logic of this paper, turning on a 
closed string tachyon leads to lower-dimensional
vacua, characterized by the decrease of $c$ along the flow \zam,
and the localization of physical excitations in space.
In the bosonic string, two dimensional string theory (\cf\ \gm) 
is a possible endpoint of the condensation.
In this stable vacuum,
the whole closed string is confined to 1+1 dimensions
(unlike the open case, where only the endpoints are confined);
all the transverse
oscillator towers of the string are frozen in the new vacuum.

The spacetime description of this drastic reduction in the number
of degrees of freedom is very similar to that of the open string.
The normalizable excitations of the string center of mass
are bound to a lower dimensional manifold.  Unlike the
open string case, the closed string oscillator degrees of freedom
don't have any normalizable excitations.

%%%%%%%%%%%%%%%%%%%%%%%%%%%%%%%%%%%%%%%%%%%%%%%%%%%%%%%%%%%%%%%%%%

%%%%%%%%%%%%%%%%%%%%%%%%%%%%%%%%%%%%%%%%%%%%%%%%%%%%%%%%%%%%%%%%%%%%%%

\vskip 1cm
\noindent{\bf Acknowledgments:} 
We are grateful to M. Douglas and P. Kraus for discussions.
This work was supported by DOE grant DE-FG02-90ER-40560
and NSF grant PHY 9901194.

\listrefs
\end